\documentclass[useAMS,usegraphicx,usenatbib]{mn2e}

\def\lap{\hbox{${_{\displaystyle<}\atop^{\displaystyle\sim}}$}}
\def\gap{\hbox{${_{\displaystyle>}\atop^{\displaystyle\sim}}$}}

\title[Radio pulsar $\ddot{\nu} - \dot{\nu}$ correlation]
{A strong $\ddot{\nu} - \dot{\nu}$ correlation in radio pulsars
with implications for torque variations}
\author[J.O. Urama, B. Link and J.M. Weisberg]{J.O. Urama,$^{1,2,3}$\thanks{E-mail:
johnson@hartrao.ac.za} B. Link$^4$ and J.M. Weisberg$^5$\\
$^{1}$Department of Physics \& Astronomy, University of Nigeria, Nsukka\\
$^{2}$Hartebeesthoek Radio Astronomy Observatory (HartRAO), P.O. Box 443,
Krugersdorp 1740, South Africa \\
$^{3}$Presently on leave at the Department of Industrial Physics, Ebonyi State University,
Abakaliki, Nigeria \\
$^4$Department of Physics, Montana State University, Bozeman, MT 59717, USA \\
$^5$Department of Physics and Astronomy, Carleton College, 1 N. College St.
Northfield, MN 55057, USA \\
  }
\begin{document}

\date{Submitted, 28 March, 2006; accepted 16 May, 2006}

\pagerange{\pageref{firstpage}--\pageref{lastpage}} \pubyear{2006}

\maketitle

\label{firstpage}

\begin{abstract}
We present an analysis of the spin-down parameters for 131 radio
pulsars for which $\ddot\nu$ has been well determined. These pulsars
have characteristic ages ranging from $10^{3} - 10^{8}$ yr and spin periods
in the range 0.4--30 s; nearly equal numbers of pulsars have $\ddot\nu>0$ as
$\ddot\nu<0$. We find a strong correlation of $\ddot\nu$ with
$\dot{\nu}$, {\em independent of the sign of} $\ddot\nu$. We
suggest that this trend can be accounted for by small, stochastic
deviations in the spin-down torque that are directly proportional (in
magnitude) to the spin-down torque.
\end{abstract}


\section{Introduction}
Isolated pulsars exhibit fascinatingly rich timing behavior. The most
dramatic variations are {\em glitches}, sudden increases in spin rate $\nu$,
that are usually accompanied by increases in the magnitude of the
spin-down rate (for a review of glitch properties see, e.g., Lyne, Shemar
\& Smith 2000, Krawczyk et al. 2003, and references therein). 
While large glitches have a well-defined signature, $(\Delta\nu ,
\Delta\dot{\nu}) = (+,-),$ microglitches exhibit all possible
signatures \citep{cordes88}. There is evidence that some isolated
pulsars undergo ``precession'' (nutation) of their spin vectors
(D'Alessandro \& McCulloch 1997; Stairs, Lyne \& Shemar 2000;
Shabanova et al. 2001; Link \& Epstein 2001; Akg\"un, Link \&
Wasserman 2006). Smaller, long-term variations in pulse phase, {\em
timing noise}, are seen in all pulsars.  Timing noise has been largely
attributed to high frequency random walks in the pulse phase, spin
frequency or spin-down rate \citep[see, for
example,][]{boynton_etal72,cordes80}. Later analysis showed that the
timing noise in some pulsars cannot be explained entirely in terms of
high-frequency random walks, but that the spin behavior is due to
discrete jumps in one or more of the spin parameters, possibly
superimposed on a high-frequency random walk \citep{cd85}. Some
pulsars show quasi-periodic phase variations over periods of years
\citep[for examples see, e.g, ][]{dr83,ale93}. The term ``timing
noise'' is perhaps a misnomer; to the extent that timing noise
represents physical processes intrinsic to the neutron star, {\em it
is part of the signal}.

Upon construction of the pulse phase history of a pulsar, $\phi(t)$,
higher order spin parameters (spin rate $\nu$ and its derivatives
$\dot\nu$ and $\ddot\nu$) are extracted from a Taylor
series fit to the observed pulse phase,
\begin{equation}
\phi(t) = \phi_0 + \nu(t-t_0) + \frac{1}{2}\dot{\nu}(t-t_0)^2 +
\frac{1}{6}\ddot{\nu}(t-t_0)^3 + ...,
\end{equation}
where $\phi_0$ is the phase at time $t_0.$ Values of $\ddot\nu$
determined in this way are typically orders of magnitude larger than
the prediction of the vacuum dipole model and often differ in sign;
such large values of $\ddot\nu$ are thought to be highly
noise-contaminated. [For example, Baykal et al. 1999, in a study of
four pulsars that show significant quadratic trends in spin rate
history, found instability in $\ddot\nu$ consistent with a noise
process]. In this connection, we note that determination of $\ddot\nu$
is a difficult task since it is not a stationary quantity; values
deduced for a given pulsar generally depend on the origin of time for
the fitting, the length of the fit and the particular data set used
(see, {\sl e.g.}, Hobbs et al. 2004, Fig. 7).  As a result, there is
considerable uncertainty in many measurements of $\ddot\nu$,
and so most quoted values should be regarded as estimates. However, to
the extent that $\ddot\nu$ is determined by an underlying noise
process that we would like to understand, reliable estimates of
$\ddot\nu$ constitute a useful statistic with which to study the
nature of timing noise in the pulsar population as a whole. [We
describe what we mean by ``reliable'' below]. The idea behind the work
we present here is to regard an estimate of $\ddot\nu$ at a given time
for a given pulsar as one outcome of the intrinsic ``noise'' process. We
consider estimates of $\ddot\nu$ in other pulsars as other
realizations of a (possibly) similar underlying process. In this way
we can quantify the strength of timing noise as a function of other
spin variables: the spin rate, its derivative and the spin-down age.

A complete characterization of timing noise has not yet been achieved,
though trends of the strength of timing noise with various spin-down
parameters have been identified. Much previous work has used
$\ddot\nu$ as the basic parameter with which to estimate the noise
strength. \citet{cordes80} quantified timing irregularities by
defining an activity parameter:
\begin{equation}
   A = \log\left[ \frac{\sigma_R(m,T)}{\sigma_R(m,T)_{Crab}}\right],
\end{equation}
where $\sigma_R(m,T)$ is the residual phase from a least-squares
polynomial fit of order $m$ over an interval of length $T$ (all
logarithms in this paper are base 10). To obtain $\sigma_R(m,T)$, the
measurement error to the rms residual is subtracted quadratically from
the observed rms residual. Cordes \& Helfand (1980) applied their
definition of $A$ for $m=2$; in an analysis of 50 pulsars, they found
evidence for correlations of $A$ with spin period, period derivative,
and spin-down age. More recently,
\citet{arz94} introduced a timing noise parameter, characterizing the
pulsar clock error caused by stochastic timing noise as
\begin{equation}
   \Delta (t) = \log\left(\frac{1}{6\nu}\left|\ddot\nu \right| t^3 \right).
\end{equation}
The statistic used by \citet{arz94} to quantify the strength of
timing noise is $\Delta_8\equiv\Delta(t=10^8\mbox{ s})$. The noise
parameter is related to the activity parameter, A, by $ A = \Delta_8 +
0.42.$ In a study of 104 pulsars, Arzoumanian et al. (1994) found a
correlation between $\Delta_8$ and $\dot p$ given roughly by
\begin{equation}
\Delta_8 = 6.6 + 0.6 \log \dot p
\end{equation}
with a characteristic spread in $\Delta_8$ of a factor of $\sim 3$,
independent of $\dot p$. 

In this Letter we demonstrate the existence of a strong correlation
between measured values of $\ddot{\nu}$ and $\dot{\nu}$. Nearly equal
numbers of pulsars have $\ddot\nu>0$ as have
$\ddot\nu<0$. Interestingly, we find that the correlations are {\em
independent of the sign of $\ddot\nu$}. Because changes in the sign of
$\ddot\nu$ with time have been seen in some pulsars, we attribute the
nearly equal numbers of pulsars with $\ddot\nu>0$ and $\ddot\nu<0$ as
reflecting stochastic variations in $\ddot\nu$ between two extremes
over some time scale. In this interpretation, our results imply a
correlation of the magnitude of timing noise (as measured by the
characteristic magnitude of $\ddot\nu$) with the spin-down torque on
the star (as measured by the observed $\dot\nu$). We suggest that the
observed trend can be accounted for by small, stochastic deviations in
the spin-down torque that are directly proportional (in magnitude) to
the spin-down torque.

\section{An analysis of spin-down parameters}

For this analysis, we have used the most reliable published values of
the measured frequency second derivative $\ddot\nu$. These include the
recently published ephemerides for 374 pulsars \citep{hobbs04} which
have individual data spans of up to 34 yr and employ a new method to
mitigate the effects of timing noise by whitening the timing
residuals; the rotational parameters, particularly $\ddot\nu$,
obtained in this way are significantly more precise than those
obtained in earlier work in the sense that they give more stable
timing solutions. While these $\ddot\nu$ values can be used in a
timing model to predict the pulse phase at an arbitrary time, Hobbs et
al. (2004) warn that they are orders of magnitude larger than the
values predicted by the dipole braking model and so may not reflect
the physics of the pulsar braking mechanism (we offer an alternative
point of view below). In general, the smaller $\vert\ddot\nu\vert$ are
less stationary than larger values, and the relative errors can exceed
100\%; we have therefore selected the subset of 127 pulsars for which
the quoted errors in $\ddot\nu$ are less than 10\%. We also include
four young pulsars whose $\ddot\nu$ have been precisely measured, so
that the Hobbs et al. (2004) de-whitening process was not
needed. These are $B0531+21$
\citep{lyne88}, $B0540-69$
\citep{cus03}, $J1119-6127$ \citep{camilo00} and $B1509-58$ \citep{kaspi94,
liv05}.  The data set shows a very large range
in pulsar spin parameters. The characteristic ages are in the range
$\sim 10^3 - 10^8$ yr and the period 0.4--30 s. We have no millisecond
pulsars in our sample. 

We consider correlations among the following spin parameters: $\nu$,
$\dot\nu$, $\ddot\nu$, the spin-down age $\tau_c\equiv
-\nu/2\dot\nu$ and $\ddot\nu_{\rm dip}\equiv 3\dot\nu^2/\nu$. 
The definitions of $\tau_c$ and $\ddot\nu_{\rm dip}$ follow from the
vacuum dipole model of secular spin evolution: $\dot\nu\propto\nu^3$. Our
results are plotted in Fig. 1, and the cross-correlation coefficients
given in Table 1. The highest correlation (91\%) is between
$\vert\ddot\nu\vert$ and $\dot\nu$, with similarly strong
correlations with $\nu_{\rm dip}$ and $\tau_c$, and a weaker
correlation with $\nu$.  The trends of Fig. 1 can be summarized by
the following fits:
\begin{equation}
\log\vert\ddot\nu\vert=-11 + \log\vert\dot\nu\vert,
\label{trend}
\end{equation}
\begin{equation}
\log\vert\ddot\nu\vert=-9.9 + 0.53 \log\ddot\nu_{\rm dip},
\end{equation}
\begin{equation}
\log\vert\ddot\nu\vert=-8.9 - 1.2 \log \tau_c. 
\end{equation}
This system of linear equations is self-consistent through the
definitions of $\tau_c$ and $\ddot\nu_{\rm dip}$. We have applied the analysis above to the full Hobbs et al. (2004)
ephemerides of nearly 400 pulsars, and we find the same trends found
above for the sub-sample of 131. We note that: 
\begin{enumerate}
\item The above trends are independent of the sign of
$\ddot\nu$. The number of pulsars with $\ddot\nu>0$ is nearly equal to
the number with $\ddot\nu<0$. Both groups follow the same trends in
magnitude of $\ddot\nu$ (Fig. 1, second panel).  
\item $\vert\ddot\nu\vert$
is always greater than or equal to $\ddot\nu_{\rm dip}$; in many cases
$\vert\ddot\nu\vert$ exceeds $\ddot\nu_{\rm dip}$ by factor of $\sim
10^6$ (Fig. 1, bottom panel). Only four pulsars fall on the
line $\ddot\nu=\ddot\nu_{\rm dip}$: B0531+21, B0540-69, J1119-6127 and
B1509-58. These pulsars appear on the upper right of Fig. 1d.
\item The correlation between $\vert\ddot\nu\vert$ and $\dot\nu$ shows
generally less scatter than found by Arzoumanian et al. (1994) for
$\Delta_8$ versus $\dot p$. 
\end{enumerate}
Aspect ii) implies very large braking indices,
$n=\nu\ddot\nu/\dot\nu^2$. The effective braking indices vary
enormously among pulsars in the sample, between $\simeq -10^{-6}$ and
$\simeq 10^6$. This in no way implies a large spin-down
torque (see below). We note that $\ddot\nu$ is
positive for the 12 pulsars with $\vert\ddot\nu\vert\gap
10^{-23}$ s$^{-3}$. While these pulsars comprise a small subset of our
total sample, they do nearly follow the dipole prediction for
$\ddot\nu$. For smaller values of $\vert\ddot\nu\vert$, there is a progression
to values of $\ddot\nu$ with both signs, accompanied by a marked deviation
from the dipole prediction. These results are consistent with the
notion that the values of $\ddot\nu$ are noise-dominated for all
but the largest values of $\vert\ddot\nu\vert$. Next, we show that
$\vert\ddot\nu\vert>>\ddot\nu_{\rm dip}$ might be due to small {\em
variations} in the spin-down torque.

\begin{figure}
\includegraphics[width=9cm]{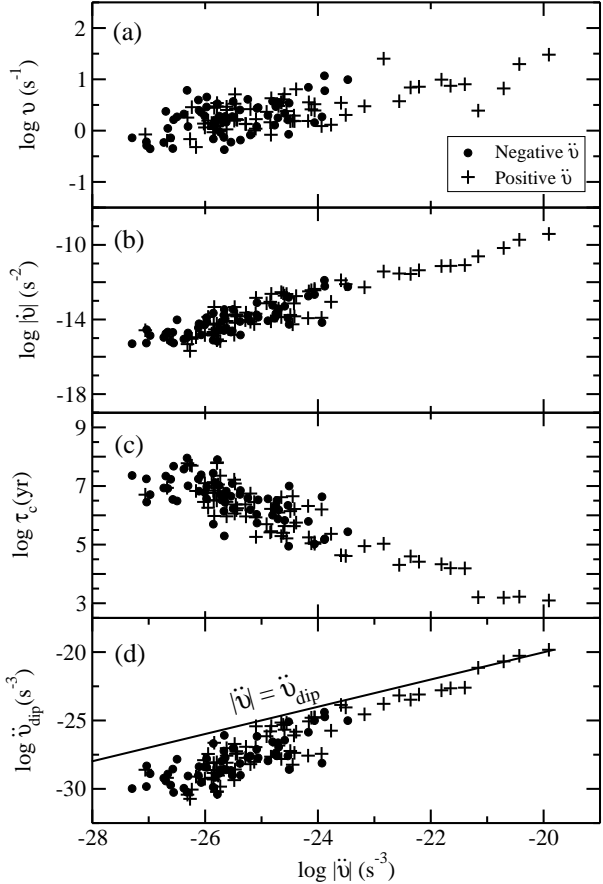}
\caption{The rotational and derived parameters versus $\ddot\nu$ for 131 pulsars.
          The (a) rotational frequency, (b) frequency derivative, (c)
          characteristic age ($ \tau_c \equiv -\nu/2\dot\nu$), and
          (d) the frequency second derivative predicted by the vacuum
          dipole model ($\ddot\nu_{\rm dip} \equiv 3\dot\nu^{2}/\nu$)
          Only four pulsars fall on the line $\ddot\nu=\ddot\nu_{\rm
          dip}$: B0531+21, B0540-69, J1119-6127 and B1509-58 (upper
right of panel d).  }
\end{figure}

\begin{table}
\caption{Cross correlation coefficients of pulsar rotational parameters. We have not included
cross-correlations among quantities that are defined in terms of each
other, e.g., $\dot\nu$ and $\tau_c$}.

\begin{tabular}{@{}lcccccc}
\hline
& $\nu$ & $\vert\dot\nu\vert$ &  $\vert\ddot\nu\vert$  & $\ddot\nu_{\rm dip}$ & $\tau_c$ \\
$\nu$ &  & 0.69 & 0.67 & & \\
$\vert\dot\nu\vert$ & & & 0.91 & & \\
$\vert\ddot\nu\vert$ & & & & 0.90 & -0.87 \\
\hline
\end{tabular}
\end{table}

\section{Discussion}

Theories of the physical origin of timing noise include 
starquakes, internal torque fluctuations, external torque fluctuations and
hybrid models that combine external torque fluctuations with
magnetospheric changes that, in turn, affect the pulsar beam
\citep[for an excellent review of timing noise models, 
see][]{ale97}. We now discuss the
possibility that the correlation between $\ddot\nu$ and
$\dot\nu$ is due to variations in the torque on the stellar
crust. External torque fluctuations could arise from, for example,
variations in currents in the pulsar magnetosphere (Cheng
1987a,b). Internal torque variations could result from vortex dynamics
in the stellar crust (e.g., Alpar et al. 1986; Jones 1990). 

First, we point out that variability in the emission region (through
movement of the region above the star, for example) cannot, by itself,
account for timing noise in pulsars with relatively large values of
$\vert\ddot\nu\vert$. A value of $\vert\ddot\nu\vert=10^{-24}$
s$^{-3}$ would lead to a phase variation of $\sim 5$ periods over 10
years. For movement of the emission region to change the phase to this
extent, the region would have to move many times around the star over
time scales of years, without producing large changes in the pulse
profile, an extremely unlikely scenario. In these pulsars, at least,
it seems that the cumulative effects of timing noise over years must
represent true variations in the rotational phase. [Short time scale
variations, however, such as phase jitter, are likely due to processes
in the emission region]. On the other hand, for pulsars with
relatively small $\vert\ddot\nu\vert$, we cannot rule out movement of
the emission region as the dominant contributor to timing noise. A
value of $\vert\ddot\nu\vert=10^{-28}$ s$^{-3}$ would lead
to phase variations of only $\lap 10^{-3}$ periods (0.4$^\circ$) over
10 years. Since this phase difference is small compared to the typical
beam width for these pulsars, the timing noise could be produced by
variations within the emission region. However, given that the
correlation we have found between $\ddot\nu$ and $\dot\nu$ holds over
the nearly eight orders of magnitude in $\ddot\nu$ that we have
considered, a single process could be at work. We henceforth assume
that timing noise represents variations in the spin rate of the star,
and that these spin variations represent variations of the torque on
the crust. We could be seeing variations in the spin-down torque on
the star, or, in principle, variations in the {\em internal} torque
exerted on the crust by the liquid interior. We consider the first
process to be more promising.

Let us write the total torque on the crust as
\begin{equation}
N_{\rm tot}=N_{\rm sd}+N_{\rm noise}, 
\end{equation}
where $N_{\rm sd}$ represents the average spin-down torque and $N_{\rm
noise}$ is a stochastic (or, at least, largely non-deterministic)
additional torque arising from fluctuations in the external torque,
the internal torque, or a combination of the two. The
observed $\dot\nu$ is
\begin{equation}
\dot\nu = \dot\nu_{\rm sd} + \dot\nu_{\rm noise}. 
\end{equation}
We expect that $\vert\dot\nu_{\rm noise}\vert$ is much smaller than
$\vert\dot\nu_{\rm sd}\vert$, while observations {\rm require} that
$\vert\ddot\nu_{\rm noise}\vert$ is much larger than
$\vert\ddot\nu_{\rm dip}\vert$. This can happen if the characteristic
time scale $T_N$ over which the noise torque varies is short compared
to the star's spin-down age. We can compare the characteristic noise
torque to the spin-down torque with a dimensionless {\em torque
parameter}, defined as: 
\begin{equation}
\tilde{N}\equiv \frac{\vert\ddot\nu\vert T_N}
{\vert\dot\nu\vert}. 
\end{equation}
The torque parameter is shown in Fig. 2 versus $\vert\dot\nu\vert$,
where we take $T_N=10$ yr as a characteristic noise time scale for
illustration (the exact value of $T_N$ is unimportant and, in any
case, varies among pulsars). We point out two features from Fig. 2:
\begin{enumerate}
\item The torque ratio $\tilde{N}$ shows no clear trend in
$\dot\nu$. {\em Torque ratios of $\tilde{N}\lap 10^{-2}$ can
account for the observed $\vert\ddot\nu\vert$($>>\ddot\nu_{\rm dip})$ in
most pulsars}.
\item Because $\tilde{N}$ is $<<1$, only small deviations from the spin-down
torque are required to produce the observed values of $\ddot\nu$.  The
magnitude of the noise torque is directly proportional to the external
{\em spin-down} torque. [The scaling with the spin down torque can be
seen in eq. (\ref{trend})].
\end{enumerate}

The stellar magnetosphere is unlikely to be completely static, and so
external torque variations are a natural explanation for our results. 
Whether or not {\em internal} torque variations applied to the crust
by the liquid interior are sufficient to account for the range in
$\ddot\nu$ values presented here, is an interesting question. 

\begin{figure}
\includegraphics[angle=-90,width=9cm]{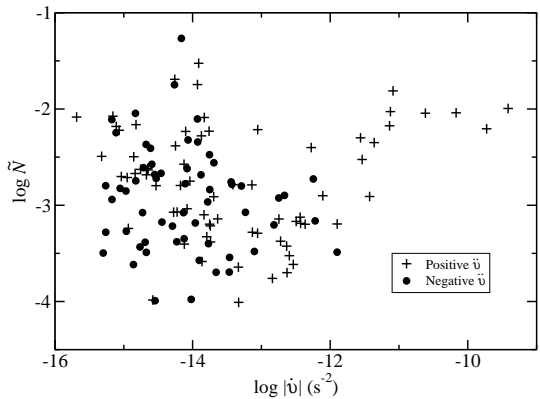}
\caption{The torque parameter 
$\tilde{N}\equiv\vert\ddot{\nu}\vert T_N/\vert\dot{\nu}\vert$ versus spin-down
rate. 
}
\end{figure}

While completing this work, we became aware of a similar analysis by
Beskin et al. (2006), who find the same correlation of $\ddot\nu$ with
$\dot\nu$ that we find. Our results are in agreement with theirs, but
our interpretation is very different. Beskin et al. (2006) attribute
the nearly equal numbers of pulsars with $\ddot\nu>0$ and $\ddot\nu<0$
to cyclic evolution of pulsar spin rates over time scales of hundreds
of years. They suggest that their assumed spin variations could
represent free-body precession over very long periods, similar to that
observed in PSR B1828-11 over a much shorter period (Stairs, Lyne \&
Shemar 2000). We emphasize, however, that the phase evolution over decades is
essentially stochastic in many cases and quasi-periodic at best in
other cases.  Only a few pulsars show strongly periodic behavior.

\section*{Acknowledgments}
Part of this work was done while JOU was visiting HartRAO and he is
very grateful to HartRAO, and the staff of Johannesburg Planetarium,
for support and hospitality. BL acknowledges the support of the
National Science Foundation under grant AST-00098728 and thanks
the University of Pisa for their hospitality, where much of this work
was completed. JMW acknowledges financial support from U.S. National
Science Foundation grant AST-0406832.

\bsp

\label{lastpage}

\end{document}